\documentclass[journal]{IEEEtran}

\pdfoutput=1

\usepackage[utf8]{inputenc}
\usepackage{graphicx}
\usepackage{amssymb}
\usepackage{amsmath}
\usepackage{wasysym}
\usepackage{siunitx}
\usepackage{color}
\usepackage{cite}

\begin{document}
\title{Fundamental Bounds on Radio Localization Precision in the Far Field}
\author
{
{Bram~J.~Dil,
Fredrik~Gustafsson,~\IEEEmembership{Fellow,~IEEE,}
and~Bernhard~J.~Hoenders}%
\thanks{B. J. Dil and F. Gustafsson are with the Department of Electrical Engineering, Linköping University, Sweden. Email: bram.j.dil@liu.se; fredrik.gustafsson@liu.se.}
\thanks{B. J. Hoenders is with the Zernike Institute for Advanced Materials at Groningen University, The Netherlands. Email: b.j.hoenders@rug.nl.}
}
\maketitle
\begin{abstract}
This paper experimentally and theoretically investigates the fundamental bounds on radio localization precision of far-field Received Signal Strength (RSS) measurements. RSS measurements are proportional to power-flow measurements time-averaged over periods long compared to the coherence time of the radiation. Our experiments are performed in a novel localization setup using 2.4GHz quasi-monochromatic radiation, which corresponds to a mean wavelength of 12.5cm. We experimentally and theoretically show that RSS measurements are cross-correlated over a minimum distance that approaches the diffraction limit, which equals half the mean wavelength of the radiation. Our experiments show that measuring RSS beyond a sampling density of one sample per half the mean wavelength does not increase localization precision, as the Root-Mean-Squared-Error (RMSE) converges asymptotically to roughly half the mean wavelength. This adds to the evidence that the diffraction limit determines (1) the lower bound on localization precision and (2) the sampling density that provides optimal localization precision. We experimentally validate the theoretical relations between Fisher information, Cramér-Rao Lower Bound (CRLB) and uncertainty, where uncertainty is lower bounded by diffraction as derived from coherence and speckle theory. When we reconcile Fisher Information with diffraction, the CRLB matches the experimental results with an accuracy of 97-98\%.
\end{abstract}
\section{Introduction} \label{sec:Introduction}
\IEEEPARstart{R}{adio} localization involves the process of obtaining physical locations using radio signals. Radio signals are exchanged between radios with known and unknown positions. Radios at known positions are called reference radios. Radios at unknown positions are called blind radios. Localization of blind radios reduces to fitting these measured radio signals to appropriate propagation models. Propagation models express the distance between two radios as a function of the measured radio signals. These measured radio signals are often modeled as deterministic radio signals with noise using a large variety of empirical statistical models \cite{HASHEMI}. Radio localization usually involves non-linear numerical optimization techniques that fit parameters in the propagation model given the joint probability distribution of the ensemble of measured radio signals. Localization precision is usually expressed as the Root Mean Squared Error (RMSE). In the field of wireless sensor networks, the estimation bounds on localization precision are often calculated by the Cramér-Rao Lower Bound (CRLB) from empirical signal models with independent noise \cite{PAT, LNSM_CRLB, CRLB_LIMIT}.

In general, localization precision depends on whether the measured radio signals contain phase information. Phase information can only be retrieved from measurements that are instantaneous on a time scale that is short relative to the oscillation period of the signals. The smallest measurable position difference depends on how well phase can be resolved. This is usually limited by the speed and noise of the electronics of the system. Less complex and less expensive localization systems are based on measurements of time-averaged power flows or Received Signal Strengths (RSS). Time-averaging is usually performed over timespans that are large compared to the coherence time of the radios, so that phase information is lost. RSS localization is an example of such less-complex systems. Hence, when determining the bounds on localization precision, it makes sense to distinguish between the time scales of the signal measurements, i.e. between instantaneous and time-averaged signal and noise processing.

Speckle theory \cite{GOODMAN2} describes the phenomenon of power-flow fluctuations due to random roughness from emitting surfaces. This theory shows that independent sources generate power-flow fluctuations in the far field that are always cross-correlated over a so-called spatial coherence region. The linear dimension of this region equals the correlation length. This correlation length has a lower bound in far-field radiation. In all practical cases of interest, the lower bound on the correlation length of this far-field coherence region is of the order of half the mean wavelength of the radiation. This lower bound is called the diffraction limit or Rayleigh criterion \cite{GOODMAN2, MANDEL}. 

It is not obvious that this lower bound on correlation length holds in our wireless sensor network setup with its characteristic small cylindrical antennas. Therefore, we derive this lower bound on the correlation length with the corresponding cross-correlation function from the Maxwell equations using the IEEE formalism described by \cite{HOOP13}. Our derivation of this lower bound on the correlation length and corresponding cross-correlation function leads to the well-known Van Cittert-Zernike theorem, known in other areas of signal processing like radar \cite{RADAR}, sonar \cite{SZABO} and optics \cite{GOODMAN3}. Our novel experimental setup validates that power-flow fluctuations are cross-correlated over a correlation length of half a wavelength by increasing the density of power-flow measurements from one to 25 power-flow measurements per wavelength.

In the field of wireless sensor networks, the estimation bounds on localization precision are usually determined by empirical signal models with independent noise over [space, time] \cite{PAT, LNSM_CRLB, CRLB_LIMIT}. Hence, the correlation length of the radiation is assumed to be infinitely small. This representation renders practical relevance as long as the inverse of the sampling rate is large compared to the correlation length, so that correlations between measurements are negligibly small. 

This paper experimentally and theoretically investigates radio localization in the far field when sufficient measurements are available to reveal the bounds on localization precision. We use Bienayme’s theorem \cite[\S 2.14]{KEEPING} to show that the ensemble of correlated power-flow measurements has an upper bound on the finite number of independent measurements over a finite measuring range. We show that this finite number of independent measurements depends on the correlation length. When we account for the finite number of independent measurements in the Fisher Information, the CRLB on localization precision deviates 2-3\% from our experimental results. We show that this approach provides practically identical results as the CRLB for signals with correlated Gaussian noise \cite[\S 3.9]{KAY93}, given the lower bound on correlation length. Hence, the lower bound on correlation length determines the upper bound on localization precision. There are a few papers that assume spatially cross-correlated Gaussian noise on power-flow measurements \cite{GUDMUNDSON, PAT08}. These papers consider cross-correlations caused by shadowing that extend over many wavelengths. \cite{PAT08} states that their cross-correlation functions do not satisfy the propagation equations. Such cross-correlations have no relation to wavelength of the carrier waves as the diffraction limit is not embedded in the propagation model like we derive.

Correlation, coherence and speckle properties of far-field radiation are governed by the spreads of Fourier pairs of wave variables that show up as variables in the classical propagation equations of electromagnetic waves. The products of these spreads of Fourier pairs express uncertainty relations in quantum mechanics, which are formulated as bandwidth relations in classical mechanics \cite{GOODMAN} and signal processing \cite{COHEN}. \cite{STAM} mathematically establishes a relationship between Fisher information, the CRLB and uncertainty. As uncertainty is lower bounded by diffraction, Fisher information is upper bounded and CRLB is lower bounded. Our experimental work reveals quantitative evidence for the validation of this theoretical work.

This paper is organized as follows. Section \ref{sec:Experimental_setup} describes the experimental setup. Section \ref{sec:theory} derives the propagation and noise models from first principles for our experimental setup for each individual member of the ensemble of measurements. Section \ref{sec:Model} describes the signal model, Maximum Likelihood Estimator (MLE) and CRLB analysis for the ensemble of measurements using the results of Section \ref{sec:theory}. Section \ref{sec:Experimental} presents the experimental results in terms of spatial correlations and upper bounds on localization precision. Finally, Section \ref{sec:discussion} provides a discussion and Section \ref{sec:conclusions} summarizes the conclusions.
\section{Experimental Setup} \label{sec:Experimental_setup}
\begin{figure} 
\centering
	\includegraphics[bb=105 270 470 555,clip=true,width=0.48\textwidth] {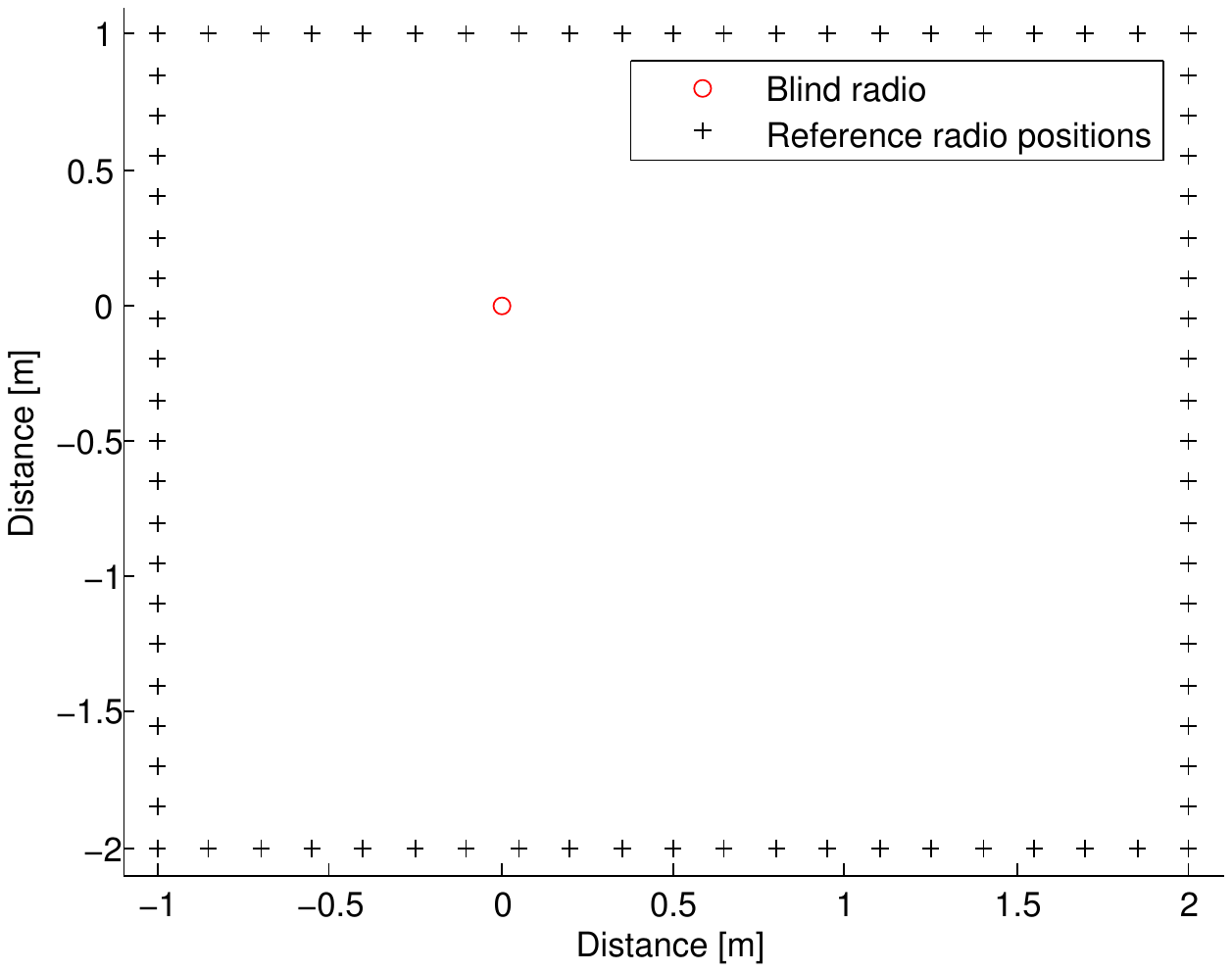}
	\caption{Measurement setup: reference radio is successively placed at known positions (crosses) on the circumference of the localization setup ($3 \times 3$m$^2$) to measure RSS to the blind radio. The blind radio is located at the origin. For illustrative purposes, this figure only shows $60$ of the $2400$ measurement positions.}
	\label{fig:Measurement_Setup}
\end{figure}

\begin{figure} [t]
\begin{center}
\includegraphics[width=0.48\textwidth] {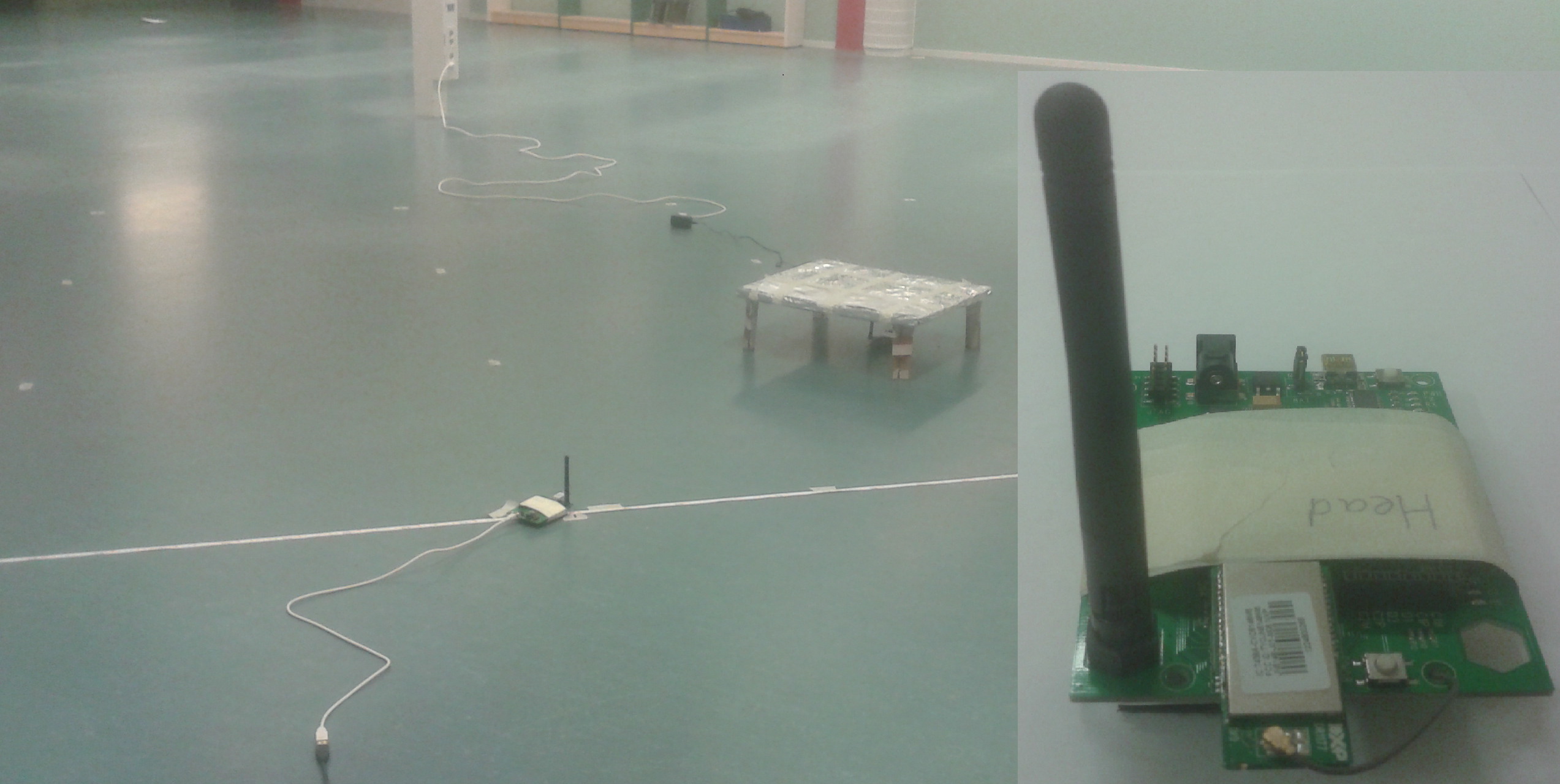}
	\caption{Photo of measurement setup: the blind radio is located underneath the aluminum table; the reference radio is located on the straight line in the picture, equidistantly positioned $0.5$cm apart. The inset on the right shows a close-up picture of the radio.}
	\label{fig:Photo}
\end{center}
\end{figure}

Fig. \ref{fig:Measurement_Setup} and \ref{fig:Photo} show the two-dimensional experimental setup. Fig. \ref{fig:Measurement_Setup} shows a square of $3\times3$m$^2$, which represents the localization surface. We distinguish between two types of radios in our measurement setup: one reference radio and one blind radio. The reference radio is successively placed at known positions and is used for estimating the position of the blind radio. The crosses represent the $2400$ manual positions of the reference radio ($x_{1},y_{1} \hdots x_{2400},y_{2400}$) and are uniformly distributed along the circumference of the square (one position every half centimeter). 

Rather than placing a two-dimensional array of reference radios inside the $3\times3$m$^2$ square, it may suffice to place a significantly smaller number on the circumference of the rectangle and get similar localization performance. Measuring field amplitudes on a circumference rather than measuring across a surface is theoretically expressed by Green's theorem as is shown in Section \ref{sec:theory}. Whether sampling power flows on circumferences instead of sampling across two-dimensional surfaces suffices has yet to be verified by experiment. This paper aims to show the practical feasibility of this novel technique, which was first proposed by \cite[\S 5.6]{DIL13}. 

The red circle represents the position of the blind radio. We place the blind radio at an unsymmetrical position, namely $(0, 0)$. We only use one blind radio and one reference radio to minimize influence of hardware differences between radios. The blind and reference radios are both main powered to minimize voltage fluctuations. The blind radio has a power amplifier and broadcasts messages with maximum power allowed by ETSI to maximize SNR. Both blind and reference radios have an external dipole antenna. The antennas have the same vertical orientation and are in Line-Of-Sight (LOS) for best reception. This implies that the polarization is vertically oriented perpendicular to the two-dimensional localization plane. The length of the antenna is half a wavelength. Its diameter is roughly one twentieth of a wavelength. We keep the relative direction of the printed circuit boards on which the antennas are mounted constant by realigning them every $25$cm. In order to minimize interference from ground reflections, we place the radios directly on the ground, so that their antennas are within one wavelength height. The ground floor consists of a reinforced concrete floor covered by industrial vinyl. We minimize interference from ceiling reflections by placing a $50\times50$cm$^2$ aluminum plate one centimeter above the blind radio antenna. The ground and the aluminum plate minimize the influence from signals in the z-direction, so that we only have to consider signals in two dimensions. All reference radio positions are in the far field of the blind radio. A photo of our setup is shown in Fig. \ref{fig:Photo}.

At each of the $2400$ reference radio positions, the reference and blind radio perform a measurement round. A measurement round consists of $500 \times 50$ repetitive multiplexed RSS measurements to investigate and quantify the measurement noise and apply CRLB analysis to this measurement noise. Each measurement round consists of $50$ measurement sets that consist of $500$ RSS measurements on an unmodulated carrier transmitted by the blind radio. Between each measurement set of $500$ RSS measurements, the reference and blind radios automatically turn on and off (recalibrate radios). Although we did not expect to find any difference from these two different forms of multiplexing RSS signals, our experiments should verify this, which they did.

Hence, a measurement round consists of $50$ measurement sets, each set consisting of $500$ repetitive multiplexed RSS measurements. Measurement rounds and measurement sets are represented by $P_{l,m,n}=P_{1,1,1} \hdots P_{2400,50,500}$. Index $l$ identifies the position of the reference radio and thus the measurement round, index $m$ identifies the $50$ measurement sets, and index $n$ identifies the $500$ individual RSS measurements in each measurement set. $P_{l,m,n}$ represents the measured power in dBm and is time-averaged over $125$\si{\micro\second} according to the radio chip specification \cite{IEEE2003}. The coherence time is $25$\si{\micro\second}, which implies a coherence length of $7.5$km. The averaging time of $125$\si{\micro\second} is a factor of five larger than the specified coherence time of the carrier wave of a typical 802.15.4 radio \cite{CC2420}. Practically, this means that these power measurements lose all phase information \cite[\S 1.5]{BROOKER}. In theory, this means that we measure the time-averaged power flow or Poynting vector as the cross-sections of the antennas are the same.

The blind radio transmits in IEEE 802.15.4 channel $15$ in order to minimize interference with Wi-Fi channels $1$ and $6$. The reference radio performs power-flow measurements in the same channel and sends the raw data to a laptop over USB, which logs the data. We use Matlab to analyze the logged data. Between each measurement round, we change the position of the reference radio by $0.5$cm and push a button to start a new measurement round. Note that $0.5$cm is well within the $\lambda / 2$ diffraction limit of half the mean wavelength. 

In summary, each measurement set takes one second; each measurement round takes roughly one minute ($50 \times 1$ seconds). The experiment consists of $2400$ measurement rounds and takes in total $40$ hours ($2400$ minutes $\approx$ $40$ hours). In practice, we spend almost three weeks in throughput time to perform this complete data collection. 
\section{Propagation and Noise Model} \label{sec:theory}
This section formulates the propagation and noise model of radio localization systems operating in the far field. This model holds for each individual member of the ensemble of $2400$ power-flow measurements over space described in Section \ref{sec:Experimental_setup}. Our propagation model is based upon EM inverse source theory, where we follow the IEEE formalism given by \cite{HOOP13}. This formalism is needed to derive the lower bound on correlation length of wireless networks using power-flow measurements on a time scale long compared to the coherence length of the hardware.
\subsection{Measurement Configuration} \label{sec:MC}
The measurement configuration of practical interest is composed of the wave-carrying medium, i.e. the atmosphere, a number of source domains and one receiver domain. Our measuring chamber has a variety of contrasting obstacles like a ceiling, walls, a ground floor, pillars, and a small aluminum table hiding the ceiling from the transmitter. Most of these secondary sources are so far away from the receiver that they are neglected in our theoretical representation. We account for the following domains:
\begin{itemize}
\item the wave-carrying medium denoted by $\mathbb{R}^{3+1}$, with electric permittivity $\epsilon_{0}>0$, magnetic permeability $\mu_{0}>0$ and EM wave speed $c_{0}=(\epsilon_{0} \mu_{0})^{-1 / 2}$. It is locally reacting, spatially invariant, time invariant and lossless. To locate positions in space, the observer employs the ordered sequence of Cartesian coordinates $\{x,y,z\} \in \mathbb{R}^3$, or $\boldsymbol{x}  \in \mathbb{R}^3$, with respect to a given origin $O$, while distances are measured through the Euclidean norm $|\boldsymbol{x}| = \sqrt{x^2+y^2+z^2}$. The fourth dimension is time and is denoted by $t$.
\item a Transmitter with bounded support $D \in \mathbb{R}^{3}$ with $D = D^{J} \cup D^{K}$. Here, $D^{J}$ denotes the spatial support carrying electric currents with volume density $J_{k}$, and $D^{K}$ denotes the spatial support carrying magnetic currents with volume density $\left[ K_{i,j} \right]^{-}$.The transmitter transmits unmodulated EM currents at a temporal frequency of $\omega_{0}/2\pi = 2.42$GHz with a temporal frequency bandwidth of $\Delta \omega_{0} / 2\pi=40$kHz \cite{CC2420}. The polarization of the EM currents is assumed to be perpendicular to the ground plane of localization space, which is assumed to be the $\{x,y\}$ plane. The electromagnetic volume current densities are usually represented by equivalent surface current densities, which physically relate to the bounded charges and currents in terms of polarization and magnetization \cite{Herczynski}.
\item Volume scatterer(s) or volume noise with bounded spatial support $D^{V} \in \mathbb{R}^{3}$ with $D^{V} = D^{V,J} \cup D^{V,K}$. Here, $D^{V,J}$ denotes the spatial support carrying electric currents with volume density $\delta J_{k}^{V}$, and $D^{V,K}$ denotes the spatial support carrying magnetic currents with volume density $\delta \left[ K_{i,j}^{V} \right]^{-}$. We assume this noise to be negligible in our setup as we identify it with the thermal noise in all contrasting media in the measuring chamber.
\item Surface scatterer(s) or surface noise with bounded spatial support $\partial D^{s} \in \mathbb{R}^{3}$ with $\partial D^{s} = \partial D^{s,J} \cup \partial D^{s,K}$. Here, $\partial D^{s,J}$ denotes the surface boundary carrying electric currents with surface density $\delta J_{k}^{s}$, and $\partial D^{s,K}$ denotes the surface boundary carrying magnetic currents with surface density $\delta \left[ K_{i,j}^{s} \right]^{-}$. The surface noise is assumed to be caused by surface roughness of the ground floor as described by \cite{GOODMAN2}.
\item a Receiver with bounded spatial support $D^{\Omega}$ in which the electric field strength $E_r$ and the magnetic field strength $\left[ H_{p,q} \right]^{-}$ are accessible to measurement through the received time-averaged power flow
\begin{equation} \label{eq:POWER}
	P_{p} = \frac{1}{T_{m}} \int_{-T_m/2}^{T_m/2} \left[ H_{p,q} \right]^{-} E_{q} dt \text{.} 
\end{equation}
The receiver measures the power flow of an unmodulated carrier with an observation time of $T_{m}=125$\si{\micro\second}, long compared to the coherence time of $25$\si{\micro\second} \cite{CC2420}. 
\end{itemize}
\subsection{Propagation Model} 
The mapping SOURCES $\Longrightarrow$ FIELD is unique if the physical condition of causality is invoked. For the received signal at the point, $\boldsymbol{x}$, of observation, the relation between the EM vector and scalar potentials, $\left\{ A_{k}, \left[ \Psi_{i,j} \right]^{-} \right\}$, of the received power flow and the real EM currents on the surfaces of the bounded sources is given by \cite[\S 6]{HOOP13}
\begin{equation} \label{eq:EH}
	\left\{ A_{k}, \left[ \Psi_{i,j} \right]^{-} \right\}(\boldsymbol{x}, t) = G(\boldsymbol{x},t) \stackrel{(\boldsymbol{x})}{*} \stackrel{(t)}{*} 
	\left\{ J_{k} , \left[ K_{i,j} \right]^{-} \right\}(\boldsymbol{x}, t) \text{,}
\end{equation}
where the spatial and temporal convolutions extend over the surfaces of the spatial-temporal supports of the pertaining bounded primary and induced sources located at $\boldsymbol{x}^{D}$ in the measurement configuration. According to Section \ref{sec:MC}, the spatial supports are the surfaces of the transmitter and of the ground floor. In two dimensions, the surfaces become circumferences of their cylindrical cross sections. Equation \eqref{eq:EH} links the extended sources to the well-known retarded potentials. In our measurement configuration, where the polarization of transmitted and received signals are directed perpendicular to the $\{x,y\}$ plane, the array of Green's functions that couples the sources to their respective potentials takes on the scalar form
\begin{equation} \label{eq:GREEN}
	G(\boldsymbol{x},t) = \frac{ \delta \left(t - \frac{\left| \boldsymbol{x} \right|}{c_{0}}\right) }{4 \pi |\boldsymbol{x}|} \text{.}
\end{equation}
In \eqref{eq:GREEN}, $\delta(t - \left| \boldsymbol{x} \right|/c_{0} )$ is the (3+1)-space-time Dirac distribution operative at $\boldsymbol{x}=\boldsymbol{0}$ and $t=0$
\begin{equation} \label{eq:dirac_delta}
	\delta \left(t - \frac{\left| \boldsymbol{x} \right|}{c_{0}} \right) 
	= \int_{-\infty}^{\infty} \exp\left[ i \omega_{0} \left(t - \frac{\left| \boldsymbol{x} \right|}{c_{0}} \right) \right] d \frac{\omega_{0}}{2 \pi} \text{.}
\end{equation}
The bandwidth, $\Delta \omega_{0}$, of temporal frequency spectrum of the hardware is usually mathematically represented by its inverse, the coherence time, $\tau_{c}$, as described by \cite[\S 10.7]{BROOKER}
\begin{equation} \label{eq:coherence_time}
	\tau_c = \frac{(8 \pi \ln 2)^{1/2}}{\Delta \omega_{0}} \text{,}
\end{equation}
where we assume a narrow Gaussian line shape relative to the carrier frequency as is the case for quasi-monochromatic radiation
\begin{equation} \label{eq:quasi_mono}
	\frac{\Delta \omega_{0}}{\omega_{0}} \ll 1 \text{.}
\end{equation}
Time and position are connected by the constant speed of light, $c_{0}$, as are angular temporal frequency and wavelength $\lambda_{0}$   
\begin{equation} \label{eq:lambda}
	\lambda_{0} = \frac{2 \pi c_{0}}{\omega_{0}} = \frac{2 \pi}{k_{0}} \text{.}
\end{equation}
To simplify the notation, we represent this time dependence of the source currents in the Dirac distribution by limiting the integration interval in \eqref{eq:dirac_delta} to the frequency bandwidth, $\Delta \omega_{0}$, 
\begin{multline} \label{eq:varying_envelope}
	\delta \left(t - \frac{\left| \boldsymbol{x} \right|}{c_{0}} \right) \cong \exp \left[ i \omega_{0} \left(t - \frac{\left| \boldsymbol{x} \right|}{c_{0}}\right) \right] \\ 
	\text{sinc} \left( \Delta \omega_{0} \left(t - \frac{\left| \boldsymbol{x} \right|}{c_{0}}\right) \right) \frac{\Delta \omega_{0}}{2 \pi}
\end{multline}
For $\Delta \omega_{0} \rightarrow 0$, the slowly varying envelope of the sinc-function approaches unity, and we end up with a plane-wave representation leading to a spherical wave in \eqref{eq:GREEN}. The sinc-function can be replaced by a normalized Gaussian or Lorentzian as is applied in \cite[\S 10.7]{BROOKER}. For our work, the only time dependence that is important is that our observation time is long compared to the coherence time so that all phase information is lost.
\subsection{Far-Field Approximation}
The convolution integral over the extended spatial supports in \eqref{eq:EH} assumes its far-field approximation when the point of observation, $\boldsymbol{x}$, is far away from any part of the extended spatial supports, the coordinates of which are denoted by $\boldsymbol{x}^{D}$, so that
\begin{equation} \label{eq:far_field}
	|\boldsymbol{x}| \gg |\boldsymbol{x}^{D}| \text{, or } 
	|\boldsymbol{x}-\boldsymbol{x}^{D}|=|\boldsymbol{x}|- \widehat{\boldsymbol{x}} \cdot \boldsymbol{x}^{D} 
	+ \mathcal{O} \left( \frac{|\boldsymbol{x}^{D}|}{|\boldsymbol{x}|} \right) \text{,}  
\end{equation}
where $\widehat{\boldsymbol{x}}$ denotes the unit vector in the direction of observation, $\boldsymbol{x}$. In the far-field approximation, the potentials in \eqref{eq:EH} are replaced by the far-field electromagnetic field strengths \cite[\S 6]{HOOP13} with the given polarization along the z-axis
\begin{equation} \label{eq:EM_far_field}
	\widetilde{E}_{z}^{\infty} (\boldsymbol{x},t)  = G^{\infty} (\boldsymbol{x},t) \stackrel{(\boldsymbol{x})}{*} \stackrel{(t)}{*} \widetilde{J}_{z}(\boldsymbol{x}, t) \text{,}
\end{equation}
where the tilde, $\widetilde{\cdot}$, denotes the sum of a macroscopic average of all primary and secondary fields including a small noise term due to the surface roughness of the ground floor
\begin{equation}
	\widetilde{E}_{z}^{\infty} (\boldsymbol{x},t) = E_{z}^{\infty} (\boldsymbol{x},t) + \delta E_{z}^{\infty} (\boldsymbol{x},t) \text{.}
\end{equation}
In \eqref{eq:EM_far_field}, the far-field Green's function is given by substituting \eqref{eq:EM_far_field} in \eqref{eq:varying_envelope} and \eqref{eq:varying_envelope} in \eqref{eq:dirac_delta}. The transverse far-field magnetic field strength, $\widetilde{H}_{t}^{\infty} (\boldsymbol{x}, t) = \left[ \widetilde{H}_{t,z}^{\infty} (\boldsymbol{x}, t) \right]^{-1}$, is directed perpendicular to the z-axis and $\widehat{\boldsymbol{x}}$ and equals the far-field electric field strength multiplied by $c_{0}\varepsilon_{0}$. In the far-field approximation, the time-averaged power flow of \eqref{eq:POWER} becomes inversely proportional to $|\boldsymbol{x}|^{-2}$ with the help of \eqref{eq:EH}, so that without any noise, \eqref{eq:POWER} reduces to
\begin{equation} \label{eq:power_ideal}
	P(\boldsymbol{x}, T_{m} \gg \tau_{c} ) = P_{|\boldsymbol{x}_{0}|} \frac{|\boldsymbol{x}_{0}|^{2}}{|\boldsymbol{x}|^{2}} \text{.} 
\end{equation}
In \eqref{eq:power_ideal}, $P$ denotes the absolute value of the time-averaged Poynting vector given by \eqref{eq:POWER}, and $P_{|\boldsymbol{x}_{0}|}$ denotes the reference power flow at reference far-field distance of, say, $|\boldsymbol{x}_{0}|=1$m. Equation \eqref{eq:power_ideal} can be read as the non-logarithmic gain model. 
\subsection{The Inverse-Source Problem}
In the inverse-source problem formulation pertaining to our measurement configuration, information is extracted from an ensemble of measured, time-averaged power-flow signals. This information reveals the nature and location of the scattering volume and surface sources in \eqref{eq:EH}. The mapping of the 
\begin{equation} \label{eq:mapping}
	\text{Observed Field } \widetilde{E}_{z}^{\infty} (\boldsymbol{x},t) \Longrightarrow \text{Generating Sources } \widetilde{J}_{z}(\boldsymbol{x}^{D}) 
\end{equation}
is known to be non-unique. A detailed analysis in this respect can be found in \cite{HOOP00}. In radio localization, an algorithm is used that is expected to lead to results with a reasonable degree of confidence. Such an algorithm is based on the iterative minimization of the norm of the mismatch in the response between an assumed propagation model for each individual measurement and an ensemble of observed power-flow signals.
\subsection{The Influence of Noise} \label{sec:Noise}
The influence of noise can be accounted for by using an input power-flow signal perturbed by an additive noise signal as is usually done in the Log-Normal Shadowing Model (LNSM). For independent, uncorrelated noise, \eqref{eq:power_ideal} reduces to the non-logarithmic gain model \cite{HASHEMI}
\begin{multline} \label{eq:Log_Normal}
	P(\boldsymbol{x}, T_{m} \gg \tau_{c}) = P_{|\boldsymbol{x}_{0}|} \frac{|\boldsymbol{x}_{0}|^{2}}{|\boldsymbol{x}|^{2}} \left( 10^{X/10} \right) \text{,}\\
	\text{for $\mathbb{E}\left[X^2\right]^{1/2}/10 \ll 1$}.
\end{multline}
In \eqref{eq:Log_Normal}, $X$ denotes the Gaussian variable of the independent zero-mean noise with standard deviation $\mathbb{E}\left[X^2\right]^{1/2}$. Under the condition that the perturbation is small, we obtain 
\begin{equation} \label{eq:linear_noise}
	10^{X/10} \cong (1+2.83 X/10) \text{, for $\mathbb{E}\left[X^2\right]^{1/2}/10 \ll 1$}.
\end{equation}
The linearization of \eqref{eq:linear_noise} is established by recursive expansion. With this linearization of the perturbed, time-averaged power flow with added noise, the nature of the correlations can be derived from \eqref{eq:POWER}, \eqref{eq:EH}, \eqref{eq:GREEN} and \eqref{eq:far_field} and then be combined with \eqref{eq:Log_Normal} and \eqref{eq:linear_noise}. We apply the rationale used by Van Cittert and Zernike.

According to this rationale, each surface element on the contrasting surfaces acts as a spatially incoherent point source represented by the Green's function of \eqref{eq:GREEN}. The contrasting surfaces observed by the receiver in the far-field in our localization setup are the surfaces of the primary and secondary sources. The primary source is the blind transmitter, and the main secondary source in our measuring chamber is the ground floor. The wave vector of the primary source is fixed in size, i.e. $k_{0} \widehat{\boldsymbol{x}}$, as the primary source is not extended and can be considered as a point source. The wave vectors $k_{0} \widehat{\boldsymbol{x}}$ of the plane wave amplitudes originating from secondary stochastic point sources induced under grazing incidence all over the ground floor are the same in size. All these scattered plane waves that are directed towards the vertical antenna of the receiver are again directed parallel to the ground floor. As the vertical receiving antenna has a circular cross-section, the projections of all these wave vectors on the planes tangential to this cross section give rise to a spatial-frequency bandwidth of $\sigma_{k} \leq k_{0}$.

The expectation value, $\mathbb{E}$, of the superposition of these plane waves in the far field at a slightly displaced position, $\boldsymbol{x} + \delta \boldsymbol{x}$, all of the same amplitude but with random phases follows from \eqref{eq:EH}, \eqref{eq:GREEN} and \eqref{eq:far_field}
\begin{align} \label{eq:Cittert_Zernike}
	&\mathbb{E} \left[ \delta E_{z}^{\infty} (\boldsymbol{x} + \delta \boldsymbol{x}) \right]	
	=\mathbb{E} \left[ G^{\infty}(\boldsymbol{x} + \delta \boldsymbol{x}) \stackrel{(\boldsymbol{x} + \delta \boldsymbol{x})}{*} \delta J_{z}^{s}(\boldsymbol{x}) \right] \nonumber \\
	&=\mathbb{E} \left[ \exp\left[ i k_{0} \widehat{\boldsymbol{x}} \cdot \delta \boldsymbol{x} \right] \delta E_{z}^{\infty}(\boldsymbol{x}) \right] \nonumber \\
	&=\mathbb{E} \left[ \exp\left[ i k_{0} \widehat{\boldsymbol{x}} \cdot \delta \boldsymbol{x} \right] \right] 
	\mathbb{E} \left[ \delta E_{z}^{\infty}(\boldsymbol{x})\right] \text{, with } \frac{|\delta \boldsymbol{x}|}{|\boldsymbol{x}|} \ll 1 \text{,}
\end{align}
where we have taken out the time dependence to simplify the notation. In \eqref{eq:Cittert_Zernike}, the third equal sign results from the spatial invariance of the propagating medium and from the fact that the variance in vertical surface roughness is statistically independent of horizontal correlations in the far-field. The ensemble average of all individual measurements over space of $\exp\left[ i k_{0} \widehat{\boldsymbol{x}} \cdot \delta \boldsymbol{x} \right]$ can be replaced by an average over the Fourier conjugated space because of the assumed wide-sense stationary (WSS) random noise process \cite[\S 3.4]{GOODMAN3}. In one dimension, the expectation value of $\exp\left[ i k_{0} \widehat{\boldsymbol{x}} \cdot \delta \boldsymbol{x} \right]$ equals the sinc-function 
\begin{multline} \label{eq:sinc}
	\mathbb{E} \left[ \exp\left[ i k_{0} \widehat{\boldsymbol{x}} \cdot \delta \boldsymbol{x} \right] \right]
	= \frac{1}{2 k_{0}} \int_{-k_{0}}^{k_{0}} \exp\left[ i k \widehat{\boldsymbol{x}} \cdot \delta \boldsymbol{x} \right] dk \\
	= \text{sinc}(k_{0} \delta x ) \text{,}
\end{multline}
and in two dimensions the jinc-function $\text{jinc}(k_{0} \delta x)$, with $\delta x = |\delta \boldsymbol{x}|$. The first-order spatial correlation function of the perturbed electric field strengths follows from \eqref{eq:Cittert_Zernike} by multiplying its deterministic form for $\widetilde{E}_{z}^{\infty} (\boldsymbol{x} + \delta \boldsymbol{x})$ by $\widetilde{E}_{z}^{*,\infty}(\boldsymbol{x})$ and by neglecting the second-order terms of the perturbed field strength. We then take the expectation value
\begin{equation} \label{eq:first_coherence}
	\frac{\mathbb{E} \left[ \widetilde{E}_{z}^{\infty} (\boldsymbol{x} + \delta \boldsymbol{x}) \widetilde{E}_{z}^{*,\infty}(\boldsymbol{x}) \right]}
	{\mathbb{E} \left[ \left| E_{z}^{\infty}(\boldsymbol{x}) \right|^{2} \right]} \cong\\ 
	1 + \frac{\mathbb{E} \left[ \delta \left| E_{z}^{\infty}(\boldsymbol{x}) \right|^{2} \right]}{\mathbb{E} \left[ \left| E_{z}^{\infty}(\boldsymbol{x}) \right|^{2} \right]} 
	\text{sinc}(k_{0} \delta x ) \text{,}
\end{equation}
where the superscript star, $^{*}$, denotes complex conjugate and the ratio in the right member denotes the normalized variance of the perturbed field. 

The sinc- and jinc-functions have their first zeros at $\delta x =\lambda_{0}/2$ and $\delta x = 0.61 \lambda_{0}$, which are called the diffraction limit or Rayleigh criterion in one and two dimensions. They can be considered as the far-field correlation functions for each individual measurement with a spatial correlation length that equals the diffraction limit. Equation \eqref{eq:first_coherence} is equivalent with the Van Cittert-Zernike theorem \cite{MANDEL}. This theorem links far-field spatial correlations of noise to diffraction. 

The diffraction limit acts as a fundamental lower bound on the spatial correlation or coherence region for time scales $T_{m} \gg \tau_{c}$, and hence, as an upper bound on the localization precision as we will see in Section \ref{sec:Model}. Equations \eqref{eq:Cittert_Zernike} - \eqref{eq:first_coherence} generally allow for extracting the phases from the EM field measurements as long as the timescales are short relative to an oscillation period $T_{m} \ll 2 \pi / \omega_{0}$. Under those time scales, the Van Cittert-Zernike theorem is equivalent with the Wiener-Khintchine theorem.

Using \eqref{eq:POWER} and \eqref{eq:first_coherence}, we extend the perturbed input signals of \eqref{eq:Log_Normal} and \eqref{eq:linear_noise} from independent to second-order spatially correlated noise to
\begin{align} \label{eq:Log_Normal_correlations}
	\frac{\mathbb{E}\left[ P(\boldsymbol{x}) P(\boldsymbol{x} + \delta \boldsymbol{x}) \right]} {\mathbb{E} \left[ P^{2}(\boldsymbol{x}) \right]}
	\cong 1+\frac{\mathbb{E}\left[ \delta \left| E_{z}^{\infty}(\boldsymbol{x}) \right|^{4} \right]} {\mathbb{E}\left[ \left| E_{z}^{\infty}(\boldsymbol{x}) \right|^{4} \right]}	
	\text{sinc}^{2}(k_{0} \delta x) \nonumber \\
	\cong	10^{\frac{\mathbb{E}\left[X^2\right] \text{sinc}^{2}(k_{0} \delta x)}{10}} \text{, for } \mathbb{E}\left[X^2\right]^{0.5}/10 \ll 1 \text{, } T_{m} \gg \tau_{c} \text{.}
\end{align}
The normalized scattering cross-sections in \eqref{eq:first_coherence} and \eqref{eq:Log_Normal_correlations} resulting from surface roughness equal the measurement variance $\mathbb{E}\left[X^2\right]$ at each individual reference-radio position. As we shall see in the next subsection, this measurement variance has a fundamental lower bound within the time scale of our measurements ($T_{m} \gg \tau_{c}$). Although the correlation function derived only holds for correlations in the close neighborhood of each point of observation, \eqref{eq:Log_Normal_correlations} can be extended to larger distances when the path losses are accounted for as is usually done in LNSM.

The correlation function adopted in the literature is an exponentially decaying function \cite{GUDMUNDSON, PAT08}. Reference \cite{PAT08} notes that the exponentially decaying correlation function does not satisfy the propagation equations of EM theory. Our correlation functions hold for each individual measurement position of the reference radio. The extension to an ensemble of measurement positions with a derivation of the measurement variance of this ensemble is given in Section \ref{sec:Model}. Section \ref{sec:Model} investigates two signal models for this extension and compares the computed measurement variances for the two correlation functions of power flows as a function of the sampling rate.
\subsection{Uncertainty and Fundamental Bounds}
Equation \eqref{eq:EM_far_field} contain two products of the Fourier pairs [angular frequency $\omega$, time $t$] and [spatial frequency $k$, position $x$]. These two sets of conjugate wave variables naturally lead to an uncertainty in physics called the Heisenberg Uncertainty principle. Hence, uncertainty is a basic property of Fourier transform pairs as described by \cite[\S 2.4.3]{PINSKY}
\begin{equation} \label{eq:Uncertainty_time}
		\sigma_{\omega} \sigma_{t} \geq 0.5 \text{,}
\end{equation}
and
\begin{equation} \label{eq:Uncertainty_space}
		\sigma_{k} \sigma_{x} \geq 0.5 \text{.}
\end{equation}
Equations \eqref{eq:Uncertainty_time} and \eqref{eq:Uncertainty_space} are the well-known uncertainty relations for classical wave variables of the Fourier pairs [energy, time] and [momentum, position]. In these Fourier pairs, energy and momentum are proportional to angular frequency and wave number, the constant of proportionality being the reduced Planck constant. Although these uncertainty relations originally stem from quantum mechanics, they fully hold in classical mechanics as bandwidth relations. In information theory, they are usually given in the time domain and are called bandwidth measurement relations as described by \cite{COHEN}. In Fourier optics, they are usually given in the space domain as described by \cite{GOODMAN}. 

Heuristically, one would expect the lower bounds on uncertainty in time, $\sigma_{t}$, and position, $\sigma_{x}$, to correspond to half an oscillation cycle, as half a cycle is the lower bound on the period of energy exchange between free-space radiation and a receiving or transmitting antenna giving a stable time-averaged Poynting vector. When one defines localization performance or precision as the inverse of the RMSE, one would expect the localization precision of the ensemble not to become infinitely large by continuously adding measurements. When the density of measurements crosses the spatial domain of spatial coherence as computed from \eqref{eq:Log_Normal_correlations}, the power-flow measurements become mutually dependent. The link between uncertainties and Fisher Information was first derived by \cite{STAM}. As we show in Section \ref{sec:Experimental}, our experiments reveal convincing evidence for this link.  
\section{Signal Model, Maximum Likelihood Estimator and Cramér-Rao Lower Bound} \label{sec:Model}
The first five subsections describe the signal model, MLE and CRLB usually applied in the field of RSS-based radio localization for independent noise \cite{PAT, LNSM_CRLB, GUS08, DIL12} and cross-correlated noise \cite{GUDMUNDSON, PAT08}. Section \ref{sec:BIENAYME} reconciles diffraction and the Nyquist sampling theorem with the CRLB using cross-correlated noise and Bienaymé's theorem. In addition, it introduces the notion of the maximum number of independent RSS measurements and relates this to Fisher Information. In Section \ref{sec:SIM}, simulations verify that the estimator is unbiased and efficient in the setup as described in Section \ref{sec:Experimental_setup}. We extend the notations introduced in Section \ref{sec:Experimental_setup} and \ref{sec:theory} that describe our measurement configuration with the bold-faced signal and estimator vectors usually employed in signal processing. The experimental setup consists of a reference radio that measures power at $2400$ positions $\boldsymbol{x}_{l}=(x_{l},y_{l}) \in \{ x_{1},y_{1} \hdots x_{2400},y_{2400} \}$ of an unmodulated carrier transmitted by the blind radio. At each position, the reference radio performs $50 \times 500 = 25,000$ repetitive multiplexed power measurements $P_{l,m,n}=P_{1,1,1} \hdots P_{2400,50,500}$. These measurements are used to estimate the blind radio position, which is located at the origin $\boldsymbol{x}^{D}=(x,y)=(0,0)$.
\subsection{Empirical Propagation Model} \label{sec:prop_model}
We adopt the empirical LNSM for modeling the power over distance decay of our RSS measurements. As the cross-sections of the blind transmitter and reference receiver are given and equal, the power as well as the power-flow measurements are assumed to satisfy the empirical LNSM \cite{HASHEMI}
\begin{equation} \label{eq:LNSM}
	P_{i} = \bar{P}(r_{i}) + X_{i} \text{,}
\end{equation}
where
\begin{equation} \label{eq:LNSM_ensemble}
	\bar{P}(r_{i}) = P_{r_{0}} - 10 \eta \log_{10}\left(\frac{r_{i}}{r_{0}}\right) \text{.}
\end{equation}
In \eqref{eq:LNSM} and \eqref{eq:LNSM_ensemble}, $i$ identifies the power-flow measurement. $\bar{P}(r_{i})$ denotes the ensemble mean of power-flow measurements at far-field distance
\begin{equation} \label{eq:r_distance}
	r_{i} = |\boldsymbol{x}_{i}-\boldsymbol{x}^{D}|
\end{equation}
in dBm. $P_{r_{0}}$ represents the power flow at reference distance $r_{0}$ in dBm. $\eta$ represents the path-loss exponent. $X_{i}$ represents the noise of the model in dB due to fading effects. $X_{i}$ follows a zero-mean Gaussian distribution with variance $\sigma_{dB}^{2}$ and is invariant with distance. Equations \eqref{eq:LNSM} and \eqref{eq:LNSM_ensemble} are equivalent with the $10 \log_{10}$ of \eqref{eq:Log_Normal}, where the path-loss exponent equals two.

One usually assumes spatially independent Gaussian noise \cite{PAT, LNSM_CRLB, GUS08, DIL12}. In \cite{GUDMUNDSON, PAT08}, the cross-correlation between power-flow measurements is independent of wavelength and is modeled with an exponentially decaying function over space by
\begin{equation} \label{eq:GUDMUNDSON}
	\rho(\delta x) = \exp \left[ \frac{-2\delta x}{\chi} \right] \text{,}
\end{equation}
where $\chi$ denotes the correlation length. Section \ref{sec:Noise} shows that the lower bound on correlation length depends on the wavelength and equals half the mean wavelength, $\chi=\lambda_{0}/2$. In Section \ref{sec:Noise}, \eqref{eq:Log_Normal_correlations} shows that the cross-correlation function of power-flow measurements takes on the form of the diffraction pattern \cite{TORALDO, GOODMAN2}
\begin{equation} \label{eq:diffraction_pattern}
	\rho(\delta x) = \text{sinc}^{2}(k_{0} \delta x) \text{.}
\end{equation}
In Section \ref{sec:CRLB}, we show the implications of using the cross-correlation function of \eqref{eq:GUDMUNDSON} and the cross-correlation function of \eqref{eq:diffraction_pattern} that satisfies the propagation model derived from first principles.
\subsection{General Framework Log-Normal Shadowing Model} 
The non-linear least squares problem assuming the LNSM with correlated noise is denoted by
\begin{equation} \label{eq:opt}
	\arg \min_{\boldsymbol{\theta}} V(\boldsymbol{\theta}) = 
	\arg \min_{\boldsymbol{\theta}} \left(\mathbf{P} - \mathbf{\bar{P}}(\mathbf{r})\right)^{T} C^{-1} \left(\mathbf{P} - \mathbf{\bar{P}}(\mathbf{r}) \right) \text{,}
\end{equation}
where $\mathbf{P}$ denotes the vector of power-flow measurements
\begin{equation} \label{eq:vec1}
	\mathbf{P} = \left[ P_{1} , \hdots , P_{n} \right] \text{,}
\end{equation}
where $\mathbf{\bar{P}}(\mathbf{r})$ denotes the vector of power-flow measurements calculated by the LNSM as expressed by \eqref{eq:LNSM_ensemble}
\begin{equation} \label{eq:vec2}
	\mathbf{\bar{P}} = \left[ \bar{P}(r_{1}), \hdots , \bar{P}(r_{n}) \right] \text{,}
\end{equation}
where $\mathbf{r}$ denotes the vector of distances between the reference radio and the blind radio
\begin{equation} \label{eq:vec3}
	\mathbf{r} = \left[ r_{1}, \hdots , r_{n} \right] \text{.} 
\end{equation}
Here, $\bar{P}(r_{i})$ denotes power flow at far-field distance $r_{i}$ as expressed by \eqref{eq:r_distance}. In \eqref{eq:opt}, $\boldsymbol{\theta}$ denotes the set of unknown parameters, where $\boldsymbol{\theta} \subseteq \left[x, y, P_{r_{0}}, \eta \right]$. $C$ denotes the covariance matrix. Usually power-flow measurements are assumed to be independent so that $C = I \sigma_{db}^{2}$, where $I$ is the identity matrix \cite{PAT, LNSM_CRLB, GUS08, DIL12}. In case of correlated noise \cite{GUDMUNDSON, PAT08}, the elements in the covariance matrix are defined by $C_{i,j} = \rho_{i,j} \sigma_{dB}^{2}$, where 
\begin{equation} \label{eq:correlation}
	\rho_{i,j}=\rho(r_{i,j}) \text{,}
\end{equation}
where
\begin{equation}
	r_{i,j} = |\boldsymbol{x}_{i}-\boldsymbol{x}_{j}|
\end{equation}
denotes the correlation distance between reference radio positions $i$ and $j$. Equation \eqref{eq:correlation} denotes the correlation coefficients as expressed by \eqref{eq:GUDMUNDSON} and \eqref{eq:diffraction_pattern}.
\subsection{Calibrate Log-Normal Shadowing Model} 
Usually, the three parameters $[P_{r_{0}}$, $\eta$, $\sigma_{dB}]$ of the LNSM are calibrated for a given localization setup \cite{PAT, LNSM_CRLB, DIL12}. The blind radio position is assumed to be known when the LNSM is calibrated. The LNSM assumes that $\sigma_{dB}$ is equal for each RSS measurement, so that $P_{r_{0}}$ and $\eta$ can be estimated independent of $\sigma_{dB}$ 
\begin{equation} \label{eq:calibrate_theta}
	[ P^{\mathrm{cal}}_{r_{0}}, \eta^{\mathrm{cal}} ]  = \arg \min_{\boldsymbol{\theta} = [ P_{r_{0}}, \eta ]} V(\boldsymbol{\theta}) \text{.}
\end{equation}
In \eqref{eq:LNSM}, $\sigma^{\mathrm{cal}}_{dB}$ is defined as the standard deviation between the measurements and the fitted LNSM using $P^{\mathrm{cal}}_{r_{0}}$ and $\eta^{\mathrm{cal}}$. Equation \eqref{eq:calibrate_theta} gives for all our power-flow measurements the best fit when the LNSM parameters are calibrated at
\begin{equation} \label{eq:calibrated}
		P^{\mathrm{cal}}_{r_0}=-16.7 \text{dBm}, \eta^{\mathrm{cal}}=3.36, \sigma^{\mathrm{cal}}_{dB}=1.68 \text{dB} \text{.}
\end{equation}
We use these calibrated LNSM parameter values to calculate the bias and efficiency of our estimator from simulations. In addition, we use these calibrated LNSM parameter values to calculate the cross-correlations in the far-field and CRLB.
\subsection{Maximum Likelihood Estimator} 
We use the MLE as proposed by \cite{GUS08, DIL12}, with the physical condition of causality invoked on $\eta$ and with boundary conditions on $(x,y)$, to estimate the blind radio position
\begin{align} \label{eq:MLE_blind_radio}
	&[ x^{\mathrm{mle}}, y^{\mathrm{mle}}, P_{r_{0}}^{\mathrm{mle}}, \eta^{\mathrm{mle}} ] = \arg \min_{\boldsymbol{\theta} = [x, y, P_{r_{0}}, \eta ]} V(\boldsymbol{\theta}) \nonumber \\
	&\text{subject to} \nonumber \\
	&\eta>0 \nonumber \\
	&-1 \leq x \leq 2 \nonumber \\
	&-2 \leq y \leq 1 \text{,}
\end{align}
Our estimator processes the noise as independent, so that $C = I \sigma_{dB}^{2}$. Section \ref{sec:SIM} shows that this estimator is unbiased and efficient in our simulation environments with uncorrelated and correlated noise.
\subsection{Cramér-Rao Lower Bound} \label{sec:CRLB}
CRLB analysis provides a lower bound on the spreads of unbiased estimators of unknown deterministic parameters. This lower bound implies that the covariance of any unbiased estimator, $\text{COV}(\widehat{\boldsymbol{\theta}})$, is bounded from below by the inverse of the Fisher Information Matrix (FIM) $F$ as given by \cite{KAY93}
\begin{equation} \label{eq:FISHER}
		\text{COV}(\widehat{\boldsymbol{\theta}}) \geq F^{-1}(\boldsymbol{\theta}) \text{.}
\end{equation}	
In \eqref{eq:FISHER}, $\boldsymbol{\theta}$ represents the set of unknown parameters, and $\widehat{\boldsymbol{\theta}}$ represents the estimator of these parameters. In case of multivariate Gaussian distributions, the elements of the FIM are given by \cite[\S 3.9]{KAY93}
\begin{multline} \label{eq:FISHER_Gaussian}
		\left[ F(\boldsymbol{\theta}) \right]_{a,b} = 
		\left[ \frac{\partial \mathbf{\bar{P}}}{\partial \theta_{a}} \right]^{T} C^{-1} \left[ \frac{\partial \mathbf{\bar{P}}}{\partial \theta_{b}} \right] + \\
		\frac{1}{2} \left[ C^{-1} \frac{\partial C}{\partial \theta_{a}} C^{-1} \frac{\partial C}{\partial \theta_{b}}\right]\text{.}
\end{multline}	
In our case, the covariance matrix is not a function of unknown parameter set $\boldsymbol{\theta}$ so that the second term equals zero. The elements of the $4 \times 4$ FIM associated with the estimator defined by \eqref{eq:MLE_blind_radio}, $\boldsymbol{\theta}=[ x, y, P_{r_{0}}, \eta ]$, are given in the independent measurement case by \cite{LNSM_CRLB} and in the correlated measurement case without nuisance parameters by \cite{PAT08}. The elements of $F$ consists of all permutations in set $\boldsymbol{\theta}$. After some algebra, $\frac{\partial \mathbf{\bar{P}}}{\partial \theta_{a}}$ reduces to
\begin{align} 
& \frac{\partial \mathbf{\bar{P}}}{\partial x_{i}} 				= -b \frac{(x-x_{i})}{r_{i}^{2}} \nonumber \\
& \frac{\partial \mathbf{\bar{P}}}{\partial y_{i}} 				= -b \frac{(y-y_{i})}{r_{i}^{2}} \nonumber \\
& \frac{\partial \mathbf{\bar{P}}}{\partial \eta_{i}}  		= - \frac{10}{\ln(10)} \ln(r_{i}) \nonumber \\
& \frac{\partial \mathbf{\bar{P}}}{\partial P_{r_{0},i}}  	= 1 \text{,} \label{eq:CRLB_MLE}
\end{align}
where
\begin{equation} 
	b = \left( \frac{10 \eta}{\ln(10)} \right) \text{.}
\end{equation}
We use \eqref{eq:CRLB_MLE} to calculate the CRLB of the MLE expressed by \eqref{eq:MLE_blind_radio} using the calibrated LNSM parameter values given by \eqref{eq:calibrated}: $[\sigma_{dB}=\sigma^{\mathrm{cal}}_{dB}, \eta=\eta^{\mathrm{cal}}]$. The CRLB on RMSE for unbiased estimators is computed from
\begin{equation} \label{eq:RMSE}
	RMSE = \sqrt{\mathbb{E}( (x^{\mathrm{mle}}-x)^2 + (y^{\mathrm{mle}}-y)^2 )} \geq \sqrt{ \mathrm{tr}(F^{-1}(\boldsymbol{\theta})) } \text{.}
\end{equation}
Here $\mathrm{tr}(\cdot)$ represents the trace of the matrix. 

\begin{figure} [t]
\centering
	\includegraphics[bb=105 270 475 560,clip=true,width=0.5\textwidth] {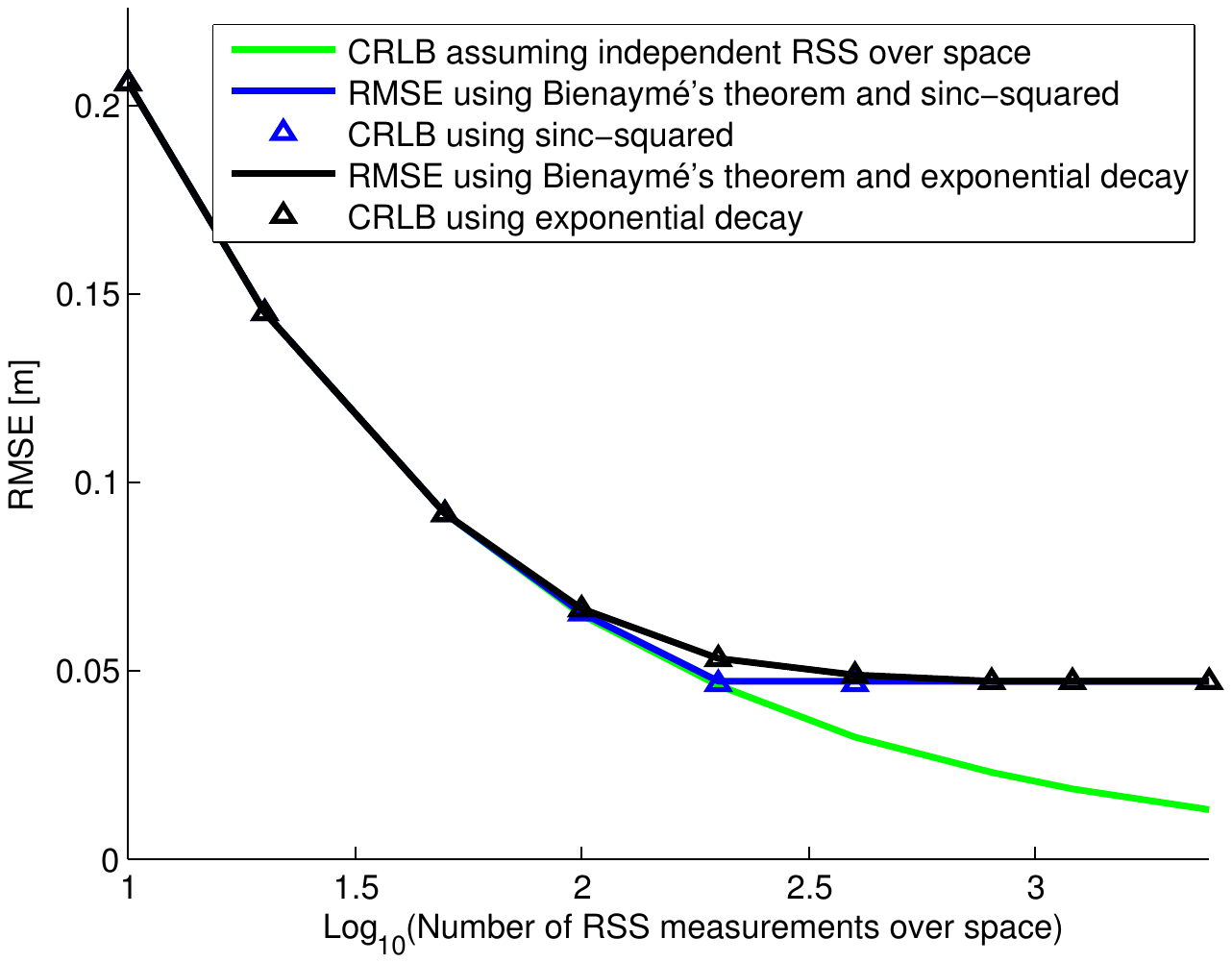}
	\caption{Theoretical RMSE calculated by CRLB as a function of the number of RSS measurements over space. The green curve shows the RMSE as a function of the number of independent RSS measurements over space. The blue curve shows the RMSE as a function of the number of RSS measurements with cross-correlations equal to the diffraction pattern as expressed by \eqref{eq:diffraction_pattern}. The blue triangles show the RMSE calculated by the corresponding CRLB as expressed by \eqref{eq:CRLB_MLE}. The covariance matrix becomes ill-conditioned when measurement density goes above a certain threshold (see text). The black curve shows the RMSE as a function of the number of RSS measurements with an exponentially decaying cross-correlation as expressed by \eqref{eq:GUDMUNDSON}. The black triangles show the RMSE calculated by the corresponding CRLB as expressed by \eqref{eq:CRLB_MLE}. These results are obtained using the same setup as described in Section \ref{sec:Experimental_setup}. We use the calibrated LNSM parameters as given by \eqref{eq:calibrated}.}
	\label{fig:CRLB}
\end{figure}

Fig. \ref{fig:CRLB} shows the lower bound on the RMSE calculated by \eqref{eq:CRLB_MLE} and \eqref{eq:RMSE} as a function of the number of RSS measurements. When we assume independent RSS measurements, $C=I \sigma_{db}^{2}$, the RMSE decreases to zero with an ever increasing number of independent RSS measurements. This is in accordance with the theoretical bound analysis presented by \cite{PAT, LNSM_CRLB, GUS08, CRLB_LIMIT}. Hence, there is no bound on localization precision with increasing sampling rate. When we account for the spatial cross-correlations between RSS measurements, $C_{i,j}=\rho_{i,j}\sigma_{dB}^{2}$, the bound on radio localization precision reveals itself when sufficient RSS measurements are available. We evaluate the CRLB with the cross-correlation function expressed by the diffraction pattern in \eqref{eq:diffraction_pattern} and the exponentially decaying cross-correlation function as expressed by \eqref{eq:GUDMUNDSON}. They basically converge to the same bound on localization precision. Apparently, the bound on localization precision depends on the bound on the correlation length rather than on the form of the cross-correlation functions.

The determinant of the covariance matrix starts to decrease when correlations start to increase, until the Fisher Information and thus localization precision stabilizes on a certain value. When the cross-correlations are equal to the diffraction pattern as expressed by \eqref{eq:diffraction_pattern}, the covariance matrix becomes singular when the sampling rate goes above a certain threshold, which equals the inverse of the diffraction limit or the Nyquist sampling rate. Hence, the multivariate Gaussian distribution becomes degenerate and the CRLB cannot be computed without regularization. 
\subsection{Diffraction, Sampling Theorem, Covariance and Fisher Information} \label{sec:BIENAYME}
This section connects diffraction to the sampling theorem, covariance and Fisher Information using Bienaymé's theorem \cite[\S 2.14]{KEEPING}. Bienaymé's theorem offers a statistical technique to estimate measurement variance of correlated signals. It states that when an estimator is a linear combination of $n$ measurements $X_1 \ldots X_n$
\begin{equation} \label{eq:linear_estimator}
	\widehat{\boldsymbol{\theta}} = \sum_{i=1}^{n} w_{i} X_{i} \text{,}
\end{equation}
the covariance of this linear combination of measurements equals
\begin{equation} \label{eq:bienayme_cov}
	\text{COV}(\widehat{\boldsymbol{\theta}}) = \sum_{i=1}^{n} w_{i}^{2} \sigma^{2}_{i} + \sum_{i=1}^{n} \sum_{j \neq i}^{n} w_{i} w_{j} \rho_{i,j} \sigma_{i} \sigma_{j}
\end{equation}	
and is equivalent to the measurement variance. In \eqref{eq:bienayme_cov}, $w_i$ is a weighting factor, $\sigma_{i}$ represents the standard deviation of measurement $i$, and $\rho_{i,j}$ represents the correlation coefficient between measurement $X_{i}$ and $X_{j}$. When all measurements have equal variance $\sigma_{i}^{2}=\sigma_{j}^{2}=\sigma^{2}$ and equal weights $w_i=w_j=1/n$, \eqref{eq:bienayme_cov} reduces to 
\begin{equation} \label{eq:bienayme_cov_equal}
	\text{COV}(\widehat{\boldsymbol{\theta}}) = \left( \frac{1}{n} + \frac{n-1}{n} \bar{\rho} \right) \sigma^{2} = \frac{1}{n_{\text{eff}}} \sigma^{2} \text{.}
\end{equation}
In \eqref{eq:bienayme_cov_equal}, $\bar{\rho}$ is the spatially averaged correlation of $n(n-1)$ measurement pairs and is given by 
\begin{equation} \label{eq:avg_rho}
	\bar{\rho} = \frac{1}{n(n-1)} \sum_{i=1}^{n} \sum_{j \neq i}^{n} \rho_{i,j}
\end{equation}
and $n_{\text{eff}}$ represents the number of measurements that effectively decreases estimator covariance and is given by
\begin{equation} \label{eq:eff_meas}
	n_{\text{eff}} = \frac{n}{1+(n-1)\bar{\rho}} \text{.}
\end{equation}
\cite{PAPOULIS} shows in the time domain that the cross-correlation of bandwidth limited fluctuations on signals with a periodic wave character is proportional to the Point-Spread-Function (PSF) of a measurement point. In the space domain, the cross-correlation of far-field power-flow measurements takes on the form expressed by the diffraction pattern in \eqref{eq:diffraction_pattern}. In addition, \cite{PAPOULIS} shows that the radius, $\delta r_{i,j}$, of the first zero 
\begin{equation} \label{eq:sampling_zero_corr}
	\rho(r_{i,j}=\lambda_{0} / 2) = 0 
\end{equation}
equals the inverse of the minimum (Nyquist) sampling rate. The sampling theorem determines the sampling rate that captures all information of a signal with finite bandwidth. For signals with a periodic wave character measured in the far field, noise is cross-correlated over the far-field spatial coherence region of the signal. This far-field coherence region assumes the far-field diffraction pattern and is equivalent to the PSF of power-flow measurements \cite{TORALDO, GOODMAN2}. 

The lower bound on the far-field region of spatially correlated signals and noise is equal to the diffraction limit \cite{GOODMAN2, MANDEL} and to the inverse of the upper bound on spatial frequencies divided by $2 \pi$, i.e. $(2k_{0}/2 \pi)^{-1}$. The finite spatial-frequency bandwidth of the far field relates diffraction to the sampling theorem \cite{TORALDO} and Whittaker-Shannon interpolation formula or Cardinal series \cite[\S 2.4]{GOODMAN}. Hence, this coherence region determines the maximum number of independent measurements over a finite measuring range. This maximum number of independent measurements is also known as the degrees of freedom \cite{TORALDO, GOODMAN}, and is given by $n_{\text{eff}}$ in \eqref{eq:eff_meas}. 

We decompose \eqref{eq:bienayme_cov_equal} into two factors 
\begin{equation} \label{eq:bienayme_two_factors}
	\text{COV}(\widehat{\boldsymbol{\theta}}) = \left( \frac{n}{n_{\text{eff}}} \right) \left( \frac{\sigma^{2}}{n} \right) \text{,}
\end{equation}	
where $\left( \frac{n}{n_{\text{eff}}} \right)$ represents the ratio of the total number of measurements and effective number of measurements, and $\left( \frac{\sigma^{2}}{n} \right)$ represents the covariance of the estimator assuming that all measurements are independent. The global spatial average of correlation coefficients, $\bar{\rho}$, as given by \eqref{eq:avg_rho} can be computed using the correlation function \eqref{eq:diffraction_pattern} for any localization setup by assuming an ever increasing sampling rate. For our localization setup of Fig. \ref{fig:Measurement_Setup}, we compute $\bar{\rho} \approx 0.0048$ processing all $2400$ RSS measurements over space. Substituting this value in \eqref{eq:eff_meas} gives $n_{\text{eff}} \approx 191$.

To account for the effects of spatial correlations on the covariance of non-linear unbiased estimators, we heuristically rewrite \eqref{eq:FISHER} into the following postulate
\begin{equation} \label{eq:bienayme_CRLB}
	\text{COV}(\widehat{\boldsymbol{\theta}}) \geq \left( \frac{n}{n_{\text{eff}}} \right) F^{-1}(\boldsymbol{\theta}) \text{,}
\end{equation}	
where $F$ is equal to the Fisher Information assuming independent measurements. Our postulate rigorously holds when the variance and thus the Fisher Information of each wave variable at each measurement point is equal, assuming that the CRLB holds as defined in \eqref{eq:FISHER}. Our postulate of \eqref{eq:bienayme_CRLB} is in agreement with the concept that the degrees of freedom of signals with a periodic wave character is independent of the estimator. Postulate \eqref{eq:bienayme_CRLB} is also in line with that CRLB efficiency is approximately maintained over nonlinear transformations if the data record is large enough \cite[\S 3.6]{KAY93}. Hence, \eqref{eq:bienayme_CRLB} is a good performance indicator when the correlation length is small relative to the linear dimensions of localization space.

Fig. \ref{fig:CRLB} shows the lower bound on the RMSE calculated by \eqref{eq:bienayme_CRLB} and \eqref{eq:RMSE} using the two cross-correlation functions expressed by \eqref{eq:GUDMUNDSON} and \eqref{eq:diffraction_pattern}. The solid curves represent the results of \eqref{eq:bienayme_CRLB}. Our approach expressed in \eqref{eq:bienayme_CRLB} differs less than 1mm from the CRLB for signals with correlated Gaussian noise \cite[\S 3.9]{KAY93} over the entire range shown by Fig. \ref{fig:CRLB}. The influence of correlated measurements on the CRLB is apparent for $n>n_{\text{eff}}$ as \eqref{eq:bienayme_CRLB} converges asymptotically to roughly half the mean wavelength for both cross-correlation functions. Hence, the correlation length and thus the finite number of independent RSS measurements determines the bound on localization precision. Note that the effective number of measurements converges to practically identical values for both cross-correlation functions, $n_{\text{eff}} \approx 191$. The sinc-squared cross-correlation function expressed by \eqref{eq:diffraction_pattern} is bandwidth limited, so that the covariance matrix becomes degenerative at roughly the Nyquist sampling rate. The exponentially decaying cross-correlation function expressed by \eqref{eq:GUDMUNDSON} is not bandwidth limited, so that the covariance matrix stays full rank. In this case, the Fisher Information decreases per individual RSS measurement with increasing sampling density until the localization precision stabilizes. When $n \approx n_{\text{eff}}$, \eqref{eq:bienayme_CRLB} provides practically the same results as the CRLB for independent measurements ($<$2mm difference in all cases). Hence, one can assume independent measurements as long as the inverse of the sampling rate is large compared to the correlation length. 

The bound on estimator covariance becomes of interest when the density of measurements surpasses the bound imposed by the spatial coherence region of far-field radiation of the transmitters. When the number of measurements goes to infinity, \eqref{eq:bienayme_cov_equal} reduces to
\begin{equation} \label{eq:COVARIANCE2}
	\lim_{n \rightarrow \infty} \text{COV}(\widehat{\boldsymbol{\theta}}) = \bar{\rho} \sigma^{2} \text{,}
\end{equation}
which does not approach zero. Hence, Bienaymé's theorem reveals the link of measurement variance to diffraction, the sampling theorem, and Fisher Information.
\subsection{Bias and Efficiency of Estimator} \label{sec:SIM}
We performed $10,000$ simulation runs using the measurement setup described in Fig. \ref{fig:Measurement_Setup} to quantify (1) the bias and (2) the efficiency of our estimator in \eqref{eq:MLE_blind_radio}. We performed two sets of simulations, one with independent noise and one with correlated noise expressed by \eqref{eq:GUDMUNDSON}. We did not perform simulations with cross-correlation function \eqref{eq:diffraction_pattern}, because the covariance matrix is singular above the Nyquist sampling rate and thus for our measurement setup. Note that these simulations do not provide an experimental validation of the model used. An estimator is unbiased if \cite{KAY93}
\begin{equation} \label{eq:unbiased}
	\mathbb{E}(\widehat{\boldsymbol{\theta}}) - \boldsymbol{\theta} = 0 \text{.}
\end{equation}
We are interested in the bias of the estimated blind radio position $(x^{mle}, y^{mle})$, so we define the bias as
\begin{equation} \label{eq:bias}
	\text{BIAS}(x^{\mathrm{mle}}, y^{\mathrm{mle}}) = \sqrt{ (\mathbb{E}(x^{\mathrm{mle}})-x)^2 + (\mathbb{E}(y^{\mathrm{mle}})-y)^2} \text{.}
\end{equation}
Simulations show that the bias is of the order of $\text{BIAS}(x^{\mathrm{mle}}, y^{\mathrm{mle}}) \cong 0.1$mm for independent and correlated noise. Estimator efficiency is defined as the difference between the covariance of the estimator and CRLB. We quantify this difference by
\begin{equation} \label{eq:efficiency}
	\text{Estimator efficiency } = |RMSE_{\text{simulated}} - \sqrt{ \mathrm{tr}(F^{-1}(\boldsymbol{\theta})) }| \text{.}
\end{equation}
Simulations show that the estimator and CRLB differs at most by $\text{Estimator efficiency}\leq 1$mm for both independent and correlated noise. Hence, our estimator in \eqref{eq:MLE_blind_radio} has a negligible bias and a high efficiency in our simulation environments with independent and correlated noise.
\section{Experimental Results} \label{sec:Experimental}
This section presents the experimental results of the two-dimensional measurement setup described in Section \ref{sec:Experimental_setup}. The first subsection presents the far-field spatial cross-correlation function of the RSS signals and estimate the spread of these spatial correlations. We then determine the spatial Fourier transform of these cross-correlations to look for an upper bound at a spatial frequency of $k=2 \pi / (\lambda_{0}/2)$, in line with the spatial resolution being bounded by the diffraction limit. Finally, we determine the global RMSE of \eqref{eq:RMSE} and show its asymptotic behavior to the diffraction limit with the increase of the density of sampling points. We compare this experimentally determined RMSE with the RMSE determined by the CRLB for independent and correlated noise. We show that the CRLB for independent noise underestimates the RMSE computed from our measurements.
%
%
%
%
%
%
%
%
%
%
%
%
%
%
%

%
%
\subsection{Spatial Correlations and Spatial Frequency Distribution} \label{sec:spatial_correlations}
\begin{figure} [t]
\centering
	\includegraphics[bb=105 270 475 560,clip=true,width=0.48\textwidth] {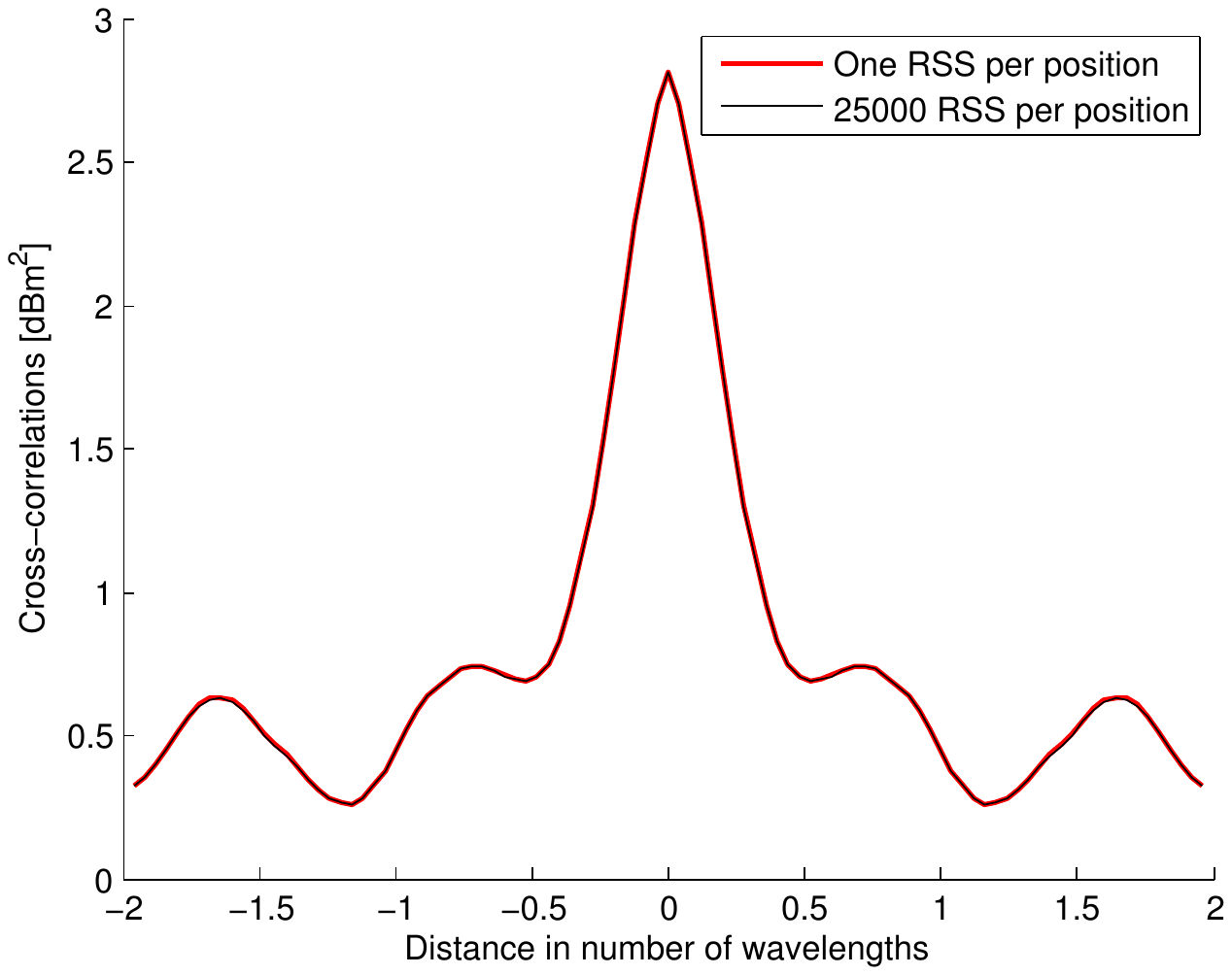}
	\caption{Spatial cross-covariance of RSS measurements as a function of correlation distance in wavelengths. The black curve shows the measured cross-correlations using all $60$ million RSS measurements. The red curve shows the measured cross-correlations using $1$ RSS measurement per reference radio position.}
	\label{fig:Coherence}
\end{figure}

Fig. \ref{fig:Coherence} shows the spatial cross-covariance function between power flows as a function of distance in wavelengths. The spatial cross-covariance function is calculated using the deviations from the LNSM expressed by \eqref{eq:LNSM} and calibrated propagation parameters expressed by \eqref{eq:calibrated}. Hence, spatial cross-covariance and thus cross-correlations are distance independent by cross correlating the deviations from the calibrated LNSM. Fig. \ref{fig:Coherence} shows that the spatial cross-correlations go to a minimum over a distance of roughly half the mean wavelength, which corresponds to the diffraction limit. The small difference between the black and red curve indicates that noise resulting from repetitive multiplexed measurements over time is negligible compared to the cross-correlated noise in the far field.

\begin{figure} [t]
\centering
	\includegraphics[bb=105 265 475 560,clip=true,width=0.48\textwidth] {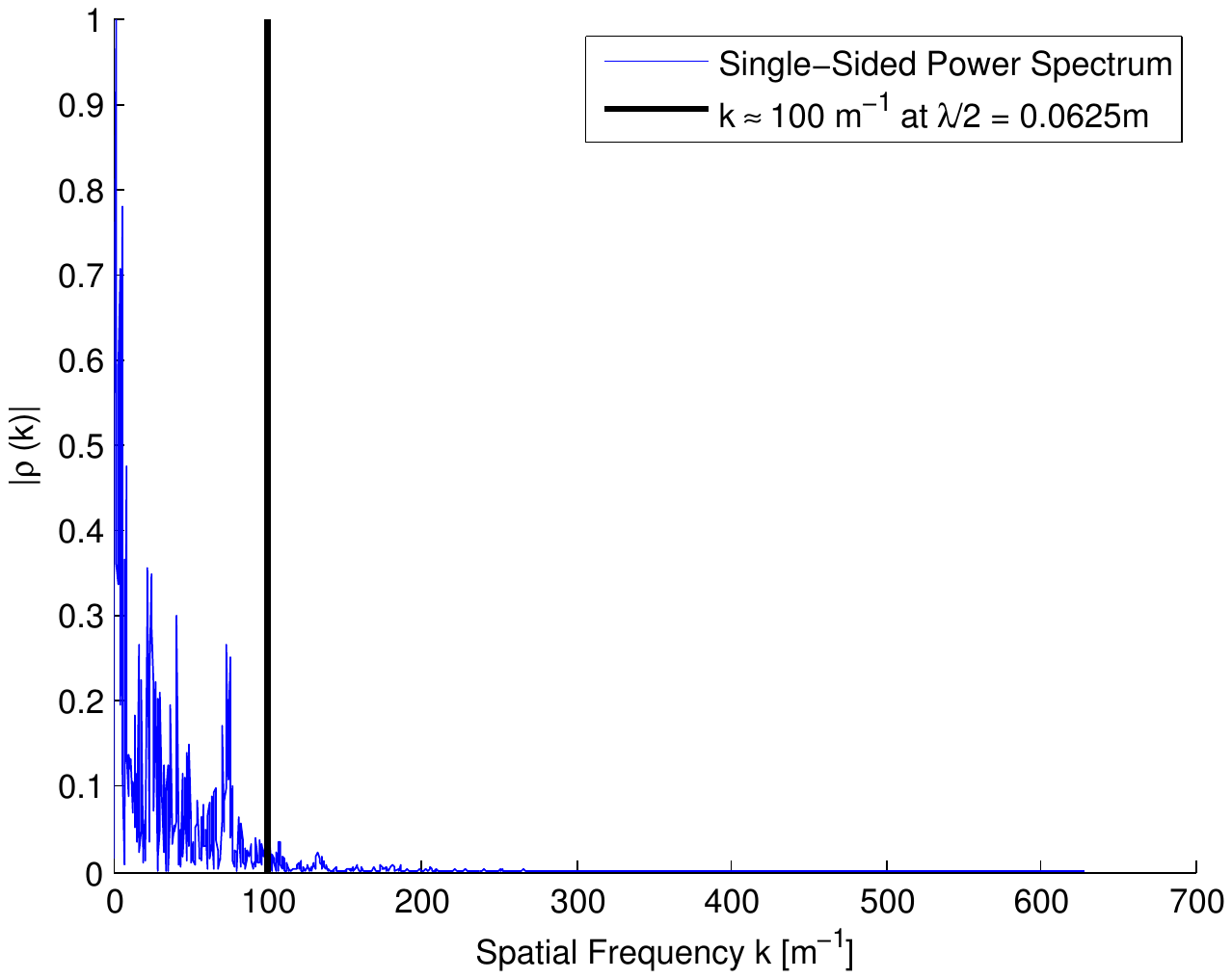}
	\caption{Measured power spectrum of cross-correlated noise in RSS signals as given by the spatial Fourier transform of the cross-covariance of Fig. \ref{fig:Coherence}. Theoretically, it is represented by tri(k), as it is the Fourier transform of \eqref{eq:diffraction_pattern}. The vertical line represents the spatial frequency that corresponds to the diffraction limit, which forms an upper bound in the spectrum.}
	\label{fig:FFT}
\end{figure}

In the case of Fisher Information and thus CRLB analysis, information is additive when measurements are independent \cite{KAY93}. One usually assumes signal models with independent noise\cite{PAT, LNSM_CRLB, CRLB_LIMIT}. Hence, localization precision increases with an ever increasing amount of independent measurements over a finite measuring range. However, when measurements are correlated, information and localization precision gain decrease with increasing correlations. Fig. \ref{fig:Coherence} shows that when space-measurement intervals become as small as the diffraction limit, measurements become spatially correlated and mutually dependent. Our measurements show that RSS signal measurements are spatially correlated over a single-sided region of roughly half the mean wavelength. It corresponds to the diffraction pattern expressed by \eqref{eq:Cittert_Zernike} derived in Section \ref{sec:Noise}. Therefore, increasing reference radio density beyond one per half a wavelength has a negligible influence on Fisher information gain and thus on localization precision. Our experimental results in the next subsection confirm this. In case of independent measurements, Fig. \ref{fig:Coherence} would show an infinitely sharp pulse (Dirac delta function).

Fig. \ref{fig:FFT} shows the measured power spectrum of cross-correlated noise in RSS signals, $|\rho(k)|$, i.e. the spatial Fourier transform of the cross-covariance of Fig. \ref{fig:Coherence}. Theoretically, it is represented by the spatial Fourier transform of \eqref{eq:diffraction_pattern}. This figure shows that the energy is mainly located in lower spatial frequencies and it diminishes over a single-sided interval of $k \leq 2\pi/(\lambda_{0}/2)$. This upper bound corresponds to the diffraction limit. The Nyquist sampling rate provides an estimate of the minimum sampling rate to fully reconstruct the power-flow signal over space without loss of information, which equals the single-sided bandwidth of our power spectrum. The vertical black line represents the spatial frequency associated with the Nyquist sampling rate, which equals $k=2\pi/(\lambda_{0}/2)$. Our experimental results in the next subsection confirm this by showing that the localization precision does not increase by sampling beyond this sampling rate. In case of independent measurements, Fig. \ref{fig:FFT} would show a uniform distribution.
\subsection{Localization Precision} \label{sec:eval_loc_precision}
\begin{figure} [t]
\centering
	\includegraphics[bb=105 270 475 560,clip=true,width=0.48\textwidth] {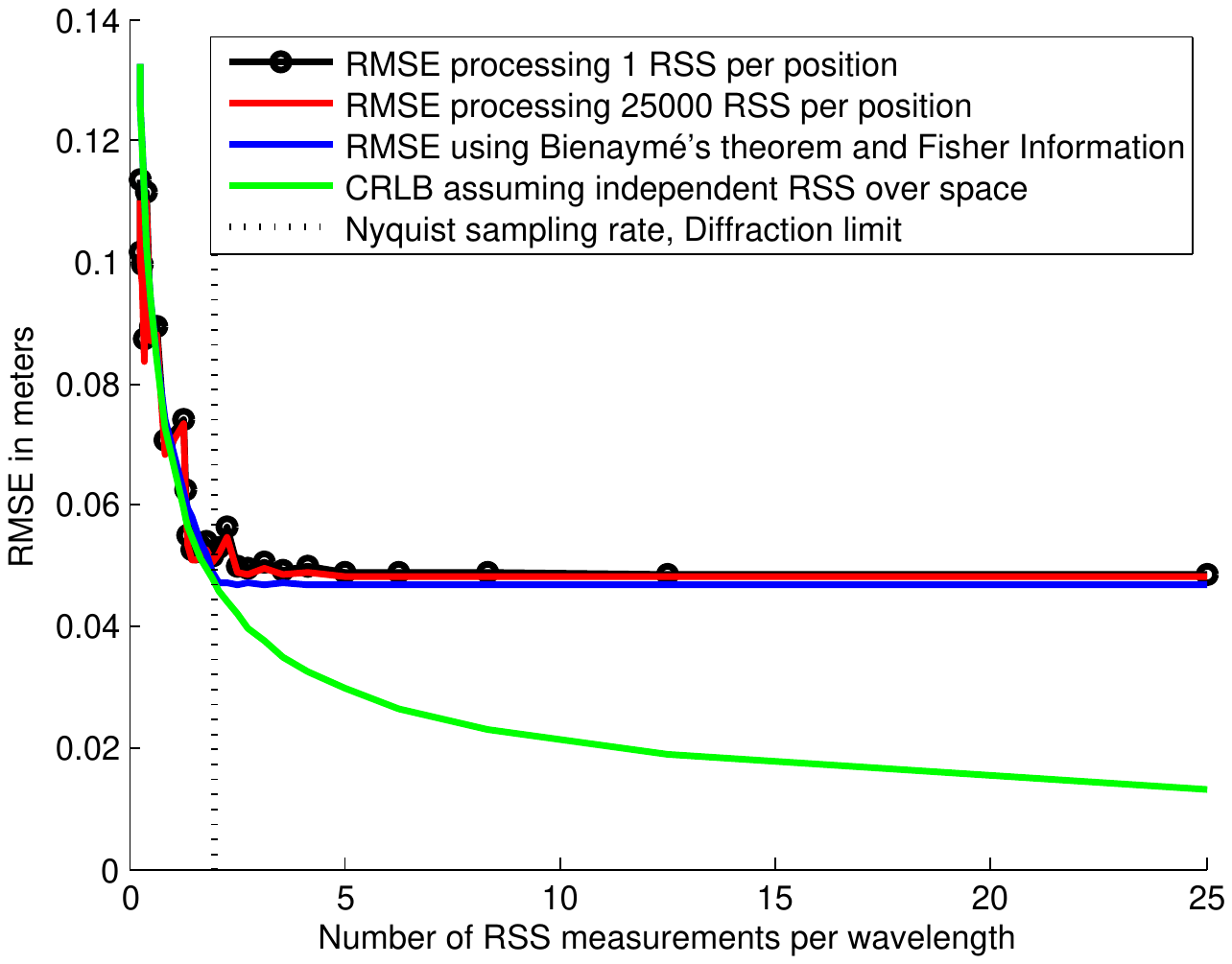}
	\caption{Experimentally determined RMSE as a function of all RSS measurements over space (red curve) with the circles indicating RMSE processing one RSS measurement per reference radio position. Estimated RMSE using Bienaymé's theorem and Fisher Information as in \eqref{eq:bienayme_CRLB} with cross-correlations equal to the diffraction pattern as expressed by \eqref{eq:diffraction_pattern} (blue curve), CRLB analysis as a function of $1$ independent RSS measurement per reference radio position (green curve).}
	\label{fig:Localization}
\end{figure}

Fig. \ref{fig:Localization} shows the experimental results of our measurement setup described in Section \ref{sec:Experimental_setup}. Localization precision is given as the inverse of RMSE, which is computed from \eqref{eq:RMSE}. 

The red curve in Fig. \ref{fig:Localization} shows the RMSE as a function of the number of measurements per wavelength. The RMSE decreases with increasing number of RSS measurements over space until sufficient measurements are available. At a critical measurement density, the bound on localization precision becomes of interest. The RMSE (red and black curves) converges asymptotically to roughly half the mean wavelength as one would expect from the diffraction limit.

The RMSE represented by the red curve is based on processing all $25000$ RSS signal measurements per reference radio position. The RMSE represented by the black curve is based on processing one RSS signal measurement per reference radio position. The negligible difference between the red and black curves shows that the number of repeated RSS signal measurements per reference radio has a negligible influence on the RMSE. 

The CRLB for independent measurements starts deviating from the RMSE (red and black curves) when the sampling rate is increased beyond one RSS signal per half the mean wavelength (see black dotted curve), as one would expect from the diffraction limit and the Nyquist sampling rate over space \cite[\S 2.4]{GOODMAN}. Spatial correlations between RSS signals increase rapidly with increasing RSS measurement density beyond one sample per half the mean wavelength (Fig. \ref{fig:Coherence}). Correlated RSS signals cannot be considered as independent. \cite{STAM} has shown that Fisher information is upper bounded by uncertainty in line with \eqref{eq:Uncertainty_space}. As $\sigma_{k}$ is upper bounded in spatial frequencies by $2k_{0}$, the spread or uncertainty in position, $\sigma_{r}$, is lower bounded. Hence its inverse is upper bounded, as is Fisher Information. Coherence and Speckle theory have shown that uncertainty is lower bounded by the diffraction limit. Hence, the CRLB at a sampling density of one sample per half the mean wavelength (vertical dotted black curve) should equal the measured bound on RMSE. We define the measured bound on RMSE as the RMSE processing all 60 million measurements. Our measurements show that the difference between this CRLB and the measured bound on RMSE is $2$-$3$\% (1mm). Hence, our experiments validate the theoretical concepts introduced by \cite{STAM}. Our experiments reveal evidence that the CRLB cannot be further decreased by increasing the number of measurements. On the other hand, at $25$ RSS signal measurements per wavelength, the RMSE is a factor of four higher than the one calculated by the CRLB for independent noise. This difference cannot be explained by the difference between the covariance of the estimator and the CRLB (see Section \ref{sec:Model}). The difference can, however, be explained by calculating the degrees of freedom of our localization setup using Bienaymé's theorem and the lower bound on correlation length. When one substitutes this lower bound in the \eqref{eq:diffraction_pattern} and substitutes \eqref{eq:diffraction_pattern} in \eqref{eq:CRLB_MLE}, one obtains the asymptotic behavior of the CRLB with correlated noise as is shown in Fig. \ref{fig:CRLB}. The measured asymptote in Fig. \ref{fig:Localization} deviates $2$-$3$\% from Bienaymé's theorem and from the CRLB for correlated noise. 

All $60$ million power measurements and Matlab files are arranged in a database at Linköping University \cite{DIL}.
\section{Discussion} \label{sec:discussion}
Our novel localization setup of using reference radios on the circumference of a two-dimensional localization area instead of setting them up in a two-dimensional array worked well. This implies that one does not have to know the phases and amplitudes on the closed surfaces around extended sources to reconstruct the positions of these extended sources. Our experiments show that it suffices to measure time-averaged power flows when a localization precision of about half a wavelength is required (in our case $6$ cm). Time averaging on a time scale that is long compared to the temporal coherence time is usually employed in RSS localization, so that phase information is lost. We expect such a setup to work well when all radio positions are in LOS. For a practical implementation in NLOS environments, we refer to \cite[\S5]{DIL13}. The propagation equations of Section \ref{sec:theory} are than applied to design an efficient setup and algorithm to locate the blind radios.  

In Section \ref{sec:theory}, we distinguished between primary and secondary cross-correlated noise in the far field. In our experimental setup, where signal levels are large compared to noise levels, it is reasonable to neglect thermal and quantum noise. Scattering from the surface roughness of the primary source cannot provide the the spatial frequency bandwidth of Fig. \ref{fig:FFT} because of the small size of the transmitting antenna. We expect that the cross-correlated noise originates from the surface roughness of the large area between the transmitter and receiver. Hence, our $2400$ time-averaged RSS measurements are correlated over space in line with the derived correlation model of \eqref{eq:diffraction_pattern} as was verified by our experiments.      

Noise can originate from a variety of sources. In our setup, the noise level of the antenna plus electronics on a typical $802.15.4$ radio is about $35$dBm below signal level as specified in \cite{CC2420}. The reproducible ripples on our RSS signals originated in part from small interference effects from undue reflections from mostly hidden metallic obstacles in the measuring chamber. Such obstacles like reinforced concrete pillars could not be removed. In a real indoor office environment, these interference effects usually dominate RSS signals. In these environments the cross-correlated noise in RSS signals is still determined by the diffraction limit, which in turn, determines the bound on localization precision. Multi-path effects and fading do not change the stochastic properties of the secondary extended sources where the noise is generated. As explained in the literature \cite{PAT08}, such models are usually clarified by ray-tracing. Ray-tracing implies the geometrical-optics approximation, so that the wave lenghth is set to zero and diffraction effects are not considered. When we set the correlation length of the exponentially decaying correlations as expressed by \cite{GUDMUNDSON} equal to the diffraction limit, the CRLB converges to practically the same bound on localization precision. It is remarkable that this bound is revealed by our simple experiment using various noise and signal models.

Our experiments reveal evidence that the diffraction limit determines the (1) bound on localization precision, (2) sampling rate that provides optimal signal reconstruction, (3) size of the coherence region and uncertainty, (4) upper bound on spatial frequencies and (5) upper bound on Fisher Information and lower bound on the CRLB. With the mathematical work of \cite{STAM}, a rigorous link between Fisher Information, CRLB and Uncertainty was established. As uncertainty has a lower bound according to coherence and speckle theory, so must Fisher Information have an upper bound, so that the CRLB has a lower bound, all when the noise processes are assumed to be Gaussian distributed. Our experiments were able to validate those theoretical concepts. This is further confirmed by applying Bienaymé's theorem. An interesting observation is that despite the fact that power flows have twice the spatial bandwidth than radiation flows determined by amplitudes, their spatial coherence regions are the same in size, as are their degrees of freedom.

Our paper describes a novel experimental and theoretical framework for estimating the lower bound on uncertainty for localization setups based on classical wave theory without any other prior knowledge. Bienaymé's theorem and existing CRLB for signals with correlated Gaussian noise reveal our postulate \eqref{eq:bienayme_CRLB}, so that the lower bound on uncertainty corresponds to the upper bound on Fisher Information. Our experimental results cannot be explained by existing propagation models with independent noise. It took almost three weeks in throughput time to perform the $60$ million measurements, generating $2400 \times 50 \times 500$ repetitive multiplexed measurements. We tried to minimize multi-path effects by avoiding the interfering influence of ground and ceiling reflections. Making sure to minimize reflections of other metallic obstacles in the measuring chamber were challenging but could be overcome. This allowed us to reveal a performance bound in theory and experiment for measurements without phase information.
\section{Conclusion} \label{sec:conclusions}
Our novel two-dimensional localization setup where we positioned the reference radios on the circumference of the localization area rather than spreading them out in a two-dimensional array over this area worked well in our LOS setup. Our measurements show that localization performance does not increase indefinitely with an ever increasing number of multiplexed RSS signals (space and time), but is limited by the spatial correlations of the far-field RSS signals. When sufficient measurements are available to minimize the influence of measurement noise on localization performance, the bound on localization precision is revealed as the region of spatially correlated far-field radiation noise. The determination of this region of spatial correlations is straightforward and can be directly calculated from RSS signals. Within this region of correlated RSS signals, assumptions of independent measurements are invalidated, so that the CRLB for independent noise underestimates the bounds on radio localization precision. This underestimation is removed by accounting for the correlations in the noise. The bound on the correlations is given by the fundamental bound on the correlation length that we derived from first principles from the propagation equations. The CRLB is linked to the uncertainty principle as measurement variance is directly related to this principle as we showed. The bounds on the precision of RSS- and TOF-based localization are expected to be equal and of the order of half a wavelength of the radiation as can be concluded from our experiments and underlying theoretical modeling. Sampling beyond the diffraction limit or the Nyquist sampling rate does not further resolve the oscillation period unless near-instantaneous measurements are performed with the a priori knowledge that signal processing is assumed to be based on non-linear mixing. Future research is aimed at the inclusion of strong interference effects such as show up in practically any indoor environment.
\section*{Acknowledgments}
The authors would like to thank Adrianus T. de Hoop, H. A. Lorentz Chair Emeritus Professor, Faculty of Electrical Engineering, Mathematics and Computer Sciences, Delft University of Technology, Delft, the Netherlands, for his comments on applying Green's theorem as a propagation model in electromagnetic theory. Secondly, they would like to thank Carlos R. Stroud, Jr. of The Institute of Optics, Rochester, NY, for sharing his insight on Fourier pairs and uncertainties. Finally, they would like to thank Gustaf Hendeby and Carsten Fritsche of Linköping University, Sweden, for their comments.
%
%
%

%
%
%
\end{document}